\documentclass[%
 reprint,
superscriptaddress,
 amsmath,amssymb,
 aps,
]{revtex4-2}
\usepackage{xcolor}
\usepackage{hyperref}
\hypersetup{breaklinks=true}
\hypersetup{hidelinks}

\usepackage{graphicx}
\usepackage{dcolumn}
\usepackage{bm}
\usepackage{xcolor}

\begin{document}
\preprint{APS/123-QED}

\title{Detection of Gravitational memory effect in LISA using triggers from ground-based detectors}

\author{Sourath Ghosh}
 \author{Alexander Weaver}%
 \affiliation{%
 Physics Department, University of Florida, USA
}%
 \author{Jose Sanjuan}
  \affiliation{%
 Department of Aerospace Engineering, Texas A\&M University, USA}%
 \author{Paul Fulda}
  \affiliation{%
 Physics Department, University of Florida, USA
}%
 \author{Guido Mueller}
\affiliation{%
 Physics Department, University of Florida, USA
}
\affiliation{Max Planck Institute for Gravitational Physics (Albert-Einstein-Institut) Hannover, Germany}%




\date{\today}
\newcommand{\SNRthousand}{14.11}
\newcommand{\SNRthousanderror}{0.93}
\newcommand{\SNRincavg}{0.50}
\newcommand{\SNRskyavg}{0.54}
\newcommand{\SNRthousandtotavg}{1.76}
\newcommand{\SNRoneeventtotavg}{0.055}
\newcommand{\SNRthreshold}{5}
\newcommand{\ERRORBARUPSNRthreshold}{5.5}
\newcommand{\ERRORBARDOWNSNRthreshold}{4.5}
\newcommand{\numeventsSNRthreshold}{8057}
\newcommand{\ROUNDnumeventsSNRthreshold}{8000}

\newcommand{\numeventsperligocatalog}{1.9}
\newcommand{\nummrepeatligocatalog}{2120.88}
\newcommand{\numyrsligocatalog}{1996.04}
\newcommand{\timeyrsligocatalog}{0.95}

\newcommand{\SNRoneLIGOthreecatalog}{0.1045}

\newcommand{\SNRaccumationrateCE}{8.69}
\newcommand{\SNRaccumationrateLIGO}{1}
\newcommand{\SNRaccumationrateligoAplus}{2.71}
\newcommand{\SNRaccumationrateligoAsharp}{4.13}
\newcommand{\SNRaccumationrateET}{18.94}
\newcommand{\SNRaccumationrateCEandET}{18.94}


\newcommand{\ROUNDnummrepeatligocatalog}{2100}
\newcommand{\timeyrsligocatalogforSNRthreshold}{1996.04}
\newcommand{\timeyrsligoAplusforSNRthreshold}{270.00}
\newcommand{\timeyrsligoAsharpforSNRthreshold}{117.27}
\newcommand{\timeyrsCEforSNRthreshold}{26.41}
\newcommand{\timeyrsETforSNRthreshold}{5.57}
\newcommand{\timeyrsCEandETforSNRthreshold}{5.57}

\newcommand{\timeyrsligocatalogforSNRthresholdAMIGO}{17.23}
\newcommand{\timeyrsligoAplusforSNRthresholdAMIGO}{2.33}
\newcommand{\timeyrsligoAsharpforSNRthresholdAMIGO}{1.01242}
\newcommand{\timeyrsCEforSNRthresholdAMIGO}{0.23}
\newcommand{\timeyrsETforSNRthresholdAMIGO}{0.048}
\newcommand{\timeyrsCEandETforSNRthresholdAMIGO}{0.048}

\newcommand{\timeyrsligocatalogforSNRthresholdALIA}{6.83}
\newcommand{\timeyrsligoAplusforSNRthresholdALIA}{0.93}
\newcommand{\timeyrsligoAsharpforSNRthresholdALIA}{0.40082}
\newcommand{\timeyrsCEforSNRthresholdALIA}{0.09}
\newcommand{\timeyrsETforSNRthresholdALIA}{0.019}
\newcommand{\timeyrsCEandETforSNRthresholdALIA}{0.019}

\newcommand{\timeyrsligocatalogforSNRthresholdFolkner}{11.14}
\newcommand{\timeyrsligoAplusforSNRthresholdFolkner}{1.50}
\newcommand{\timeyrsligoAsharpforSNRthresholdFolkner}{0.65458}
\newcommand{\timeyrsCEforSNRthresholdFolkner}{0.14}
\newcommand{\timeyrsETforSNRthresholdFolkner}{0.031}
\newcommand{\timeyrsCEandETforSNRthresholdFolkner}{ 0.031}


\newcommand{\ROUNDtimeyrsligocatalogforSNRthreshold}{2000}
\newcommand{\ROUNDtimeyrsligoAplusforSNRthreshold}{270}
\newcommand{\ROUNDtimeyrsligoAsharpforSNRthreshold}{120}
\newcommand{\ROUNDtimeyrsCEforSNRthreshold}{26}
\newcommand{\ROUNDtimeyrsETforSNRthreshold}{6}
\newcommand{\ROUNDtimeyrsCEandETforSNRthreshold}{6}

\newcommand{\ROUNDtimeyrsligocatalogforSNRthresholdAMIGO}{17}
\newcommand{\ROUNDtimeyrsligoAplusforSNRthresholdAMIGO}{2.3}
\newcommand{\ROUNDtimeyrsligoAsharpforSNRthresholdAMIGO}{1.0}
\newcommand{\ROUNDtimeyrsCEforSNRthresholdAMIGO}{2.3\times 10^{-1}}
\newcommand{\ROUNDtimeyrsETforSNRthresholdAMIGO}{4.8\times 10^{-2}}
\newcommand{\ROUNDtimeyrsCEandETforSNRthresholdAMIGO}{4.8\times 10^{-2}}

\newcommand{\ROUNDtimeyrsligocatalogforSNRthresholdALIA}{7}
\newcommand{\ROUNDtimeyrsligoAplusforSNRthresholdALIA}{0.9}
\newcommand{\ROUNDtimeyrsligoAsharpforSNRthresholdALIA}{0.4}
\newcommand{\ROUNDtimeyrsCEforSNRthresholdALIA}{9\times 10^{-2}}
\newcommand{\ROUNDtimeyrsETforSNRthresholdALIA}{1.9\times 10^{-2}}
\newcommand{\ROUNDtimeyrsCEandETforSNRthresholdALIA}{1.9\times 10^{-2}}

\newcommand{\ROUNDtimeyrsligocatalogforSNRthresholdFolkner}{11}
\newcommand{\ROUNDtimeyrsligoAplusforSNRthresholdFolkner}{1.5}
\newcommand{\ROUNDtimeyrsligoAsharpforSNRthresholdFolkner}{0.7}
\newcommand{\ROUNDtimeyrsCEforSNRthresholdFolkner}{1.4\times 10^{-1}}
\newcommand{\ROUNDtimeyrsETforSNRthresholdFolkner}{3.1\times 10^{-2}}
\newcommand{\ROUNDtimeyrsCEandETforSNRthresholdFolkner}{3.1\times 10^{-2}}

\newcommand{\ERRORBOUNDUPtimeyrsligocatalogforSNRthresholdERRORBOUNDUP}{+419.17}
\newcommand{\ERRORBOUNDUPtimeyrsligoAplusforSNRthresholdERRORBOUNDUP}{+56.70}
\newcommand{\ERRORBOUNDUPtimeyrsCEforSNRthresholdERRORBOUNDUP}{+5.44}
\newcommand{\ERRORBOUNDUPtimeyrsETforSNRthresholdERRORBOUNDUP}{+1.17}
\newcommand{\ERRORBOUNDUPtimeyrsCEandETforSNRthresholdERRORBOUNDUP}{+1.17}

\newcommand{\ERRORBOUNDUPtimeyrsligocatalogforSNRthresholdAMIGOERRORBOUNDUP}{+3.62}
\newcommand{\ERRORBOUNDUPtimeyrsligoAplusforSNRthresholdAMIGOERRORBOUNDUP}{+0.50}
\newcommand{\ERRORBOUNDUPtimeyrsCEforSNRthresholdAMIGOERRORBOUNDUP}{+0.05}
\newcommand{\ERRORBOUNDUPtimeyrsETforSNRthresholdAMIGOERRORBOUNDUP}{+0.01}
\newcommand{\ERRORBOUNDUPtimeyrsCEandETforSNRthresholdAMIGOERRORBOUNDUP}{+0.01}

\newcommand{\ERRORBOUNDUPtimeyrsligocatalogforSNRthresholdALIAERRORBOUNDUP}{+1.44}
\newcommand{\ERRORBOUNDUPtimeyrsligoAplusforSNRthresholdALIAERRORBOUNDUP}{+0.20}
\newcommand{\ERRORBOUNDUPtimeyrsCEforSNRthresholdALIAERRORBOUNDUP}{+0.02}
\newcommand{\ERRORBOUNDUPtimeyrsETforSNRthresholdALIAERRORBOUNDUP}{+0.004}
\newcommand{\ERRORBOUNDUPtimeyrsCEandETforSNRthresholdALIAERRORBOUNDUP}{+0.004}

\newcommand{\ERRORBOUNDUPtimeyrsligocatalogforSNRthresholdFolknerERRORBOUNDUP}{+2.34}
\newcommand{\ERRORBOUNDUPtimeyrsligoAplusforSNRthresholdFolknerERRORBOUNDUP}{+0.32}
\newcommand{\ERRORBOUNDUPtimeyrsCEforSNRthresholdFolknerERRORBOUNDUP}{+0.03}
\newcommand{\ERRORBOUNDUPtimeyrsETforSNRthresholdFolknerERRORBOUNDUP}{+0.007}
\newcommand{\ERRORBOUNDUPtimeyrsCEandETforSNRthresholdFolknerERRORBOUNDUP}{+0.007}

\newcommand{\ERRORBOUNDDOWNtimeyrsligocatalogforSNRthresholdERRORBOUNDDOWN}{-379.25}
\newcommand{\ERRORBOUNDDOWNtimeyrsligoAplusforSNRthresholdERRORBOUNDDOWN}{-51.30}
\newcommand{\ERRORBOUNDDOWNtimeyrsCEforSNRthresholdERRORBOUNDDOWN}{-4.92}
\newcommand{\ERRORBOUNDDOWNtimeyrsETforSNRthresholdERRORBOUNDDOWN}{-1.05}
\newcommand{\ERRORBOUNDDOWNtimeyrsCEandETforSNRthresholdERRORBOUNDDOWN}{-1.05}

\newcommand{\ERRORBOUNDDOWNtimeyrsligocatalogforSNRthresholdAMIGOERRORBOUNDDOWN}{-3.30}
\newcommand{\ERRORBOUNDDOWNtimeyrsligoAplusforSNRthresholdAMIGOERRORBOUNDDOWN}{-0.44}
\newcommand{\ERRORBOUNDDOWNtimeyrsCEforSNRthresholdAMIGOERRORBOUNDDOWN}{-0.04}
\newcommand{\ERRORBOUNDDOWNtimeyrsETforSNRthresholdAMIGOERRORBOUNDDOWN}{-0.01}
\newcommand{\ERRORBOUNDDOWNtimeyrsCEandETforSNRthresholdAMIGOERRORBOUNDDOWN}{-0.01}

\newcommand{\ERRORBOUNDDOWNtimeyrsligocatalogforSNRthresholdALIAERRORBOUNDDOWN}{-1.30}
\newcommand{\ERRORBOUNDDOWNtimeyrsligoAplusforSNRthresholdALIAERRORBOUNDDOWN}{-0.18}
\newcommand{\ERRORBOUNDDOWNtimeyrsCEforSNRthresholdALIAERRORBOUNDDOWN}{-0.02}
\newcommand{\ERRORBOUNDDOWNtimeyrsETforSNRthresholdALIAERRORBOUNDDOWN}{-0.004}
\newcommand{\ERRORBOUNDDOWNtimeyrsCEandETforSNRthresholdALIAERRORBOUNDDOWN}{-0.004}

\newcommand{\ERRORBOUNDDOWNtimeyrsligocatalogforSNRthresholdFolknerERRORBOUNDDOWN}{-2.12}
\newcommand{\ERRORBOUNDDOWNtimeyrsligoAplusforSNRthresholdFolknerERRORBOUNDDOWN}{-0.30}
\newcommand{\ERRORBOUNDDOWNtimeyrsCEforSNRthresholdFolknerERRORBOUNDDOWN}{-0.03}
\newcommand{\ERRORBOUNDDOWNtimeyrsETforSNRthresholdFolknerERRORBOUNDDOWN}{-0.006}
\newcommand{\ERRORBOUNDDOWNtimeyrsCEandETforSNRthresholdFolknerERRORBOUNDDOWN}{ -0.006}

\newcommand{\timeyrsligocatalogforSNRthresholdone}{79.85}
\newcommand{\timeyrsligoAplusforSNRthresholdone}{10.80}
\newcommand{\timeyrsCEforSNRthresholdone}{1.06}
\newcommand{\timeyrsETforSNRthresholdone}{0.23}
\newcommand{\timeyrsCEandETforSNRthresholdone}{0.23}

\newcommand{\timeyrsligocatalogforSNRthresholdtwoshotnoisethree}{114.94}
\newcommand{\timeyrsligoAplusforSNRthresholdtwoshotnoisethree}{15.55}
\newcommand{\timeyrsCEforSNRthresholdtwoshotnoisethree}{1.53}
\newcommand{\timeyrsETforSNRthresholdtwoshotnoisethree}{0.33}
\newcommand{\timeyrsCEandETforSNRthresholdtwoshotnoisethree}{0.33}

\begin{abstract}
The LIGO-Virgo-Kagra (LVK) collaboration has detected gravitational waves from 90 Compact Binary Coalescences. In addition to fortifying the linearized theory of General Relativity (GR), the statistical ensemble of detections also provides prospects of detecting nonlinear effects predicted by GR, one such prediction being the nonlinear gravitational memory effect. For detected stellar and intermediate mass compact binaries, the  induced strain from the memory effect is one or two orders below the detector noise background.  Additionally, since most of the energy is radiated at merger the strain induced by the memory effect resembles a step function at the merger time. These facts motivate the idea of coherently stacking up data streams from recorded GW events at these merger times so that the cumulative memory strain is detected with a sufficient SNR. GW detectors essentially record the integrated strain response at time scales of the round trip light travel time,  making future space-based long arm interferometers like LISA ideal for detecting the memory effect at low frequencies.  In this paper, we propose a method that uses the event catalog of ground-based detectors and searches for corresponding memory strains in the LISA data stream. Given LVK's O3 science run catalog, we use scaling arguments and assumptions on the source population models to estimate the run time required for LISA to accumulate a memory SNR of $\SNRthreshold$, using triggers from current and future ground-based detectors. Finally, we extend these calculations for using beyond LISA missions like ALIA, AMIGO, and Folkner to detect the gravitational memory effect. The results for LISA indicate a possible detection of the memory effect within the 10 year LISA mission lifetime and the corresponding results for beyond LISA missions are even more promising.
\end{abstract}
\maketitle

\newcommand{\lumdist}{D_{\rm{L}}}
\newcommand{\comdist}{D_{\rm{C}}}

\section{Introduction}
\begin{figure}[ht]
    \includegraphics[scale=0.15]{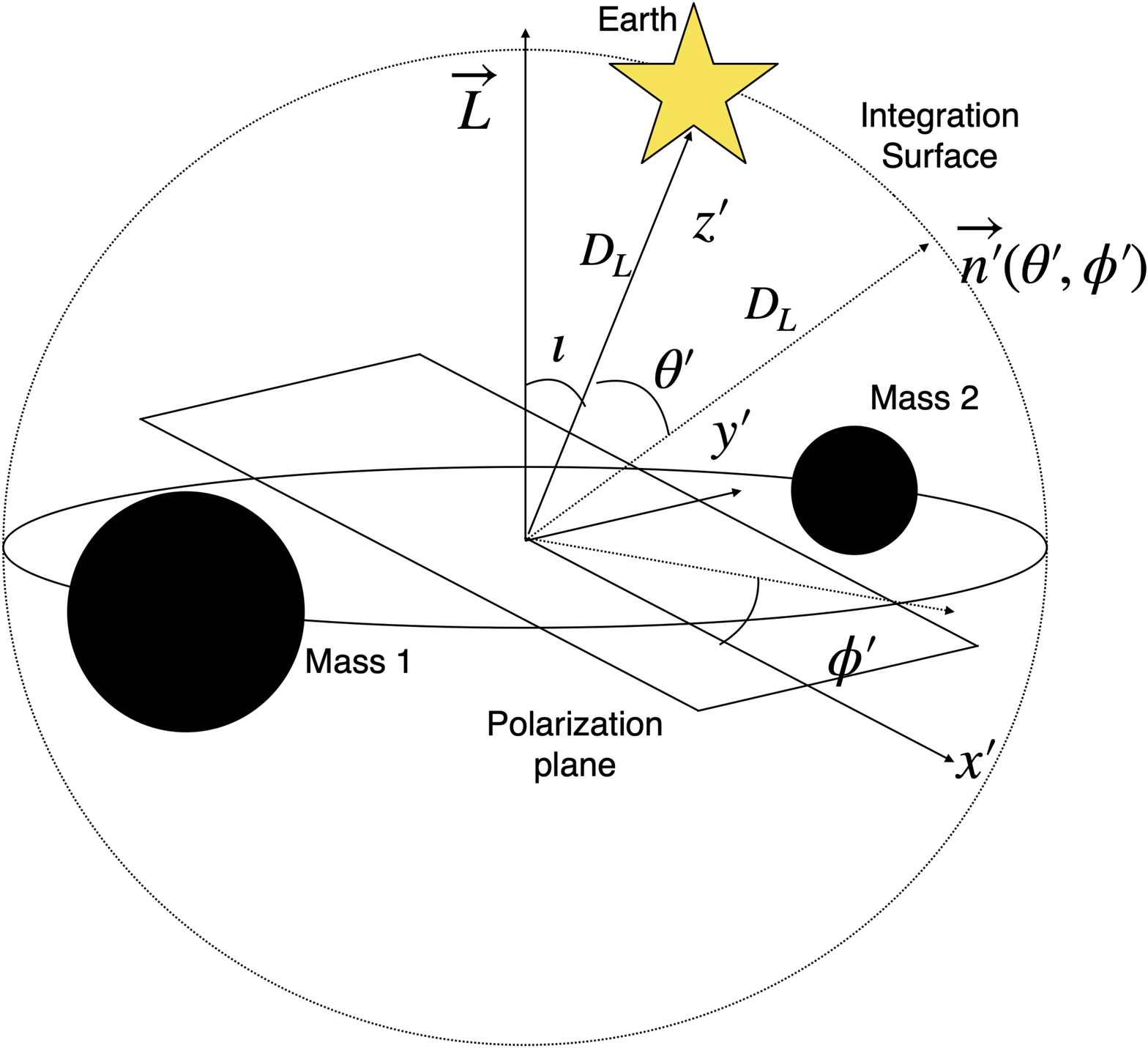}
        \caption{Source frame coordinate system for compact binary sources. $z'$ axis is fixed as the line of sight direction, and $\iota$ is the inclination angle, i.e., the angle between the source angular momentum $\vec{L}$ and the line of sight. Furthermore $x'$ axis is defined to be co-planar to $z'$ and $\vec{L}$, while $y'$ axis is defined so that $x',y',z'$ form an orthogonal right-handed coordinate system}
    \label{fig:Coordinates_source}
\end{figure}

\begin{figure}[ht]
        \includegraphics[scale=0.35]{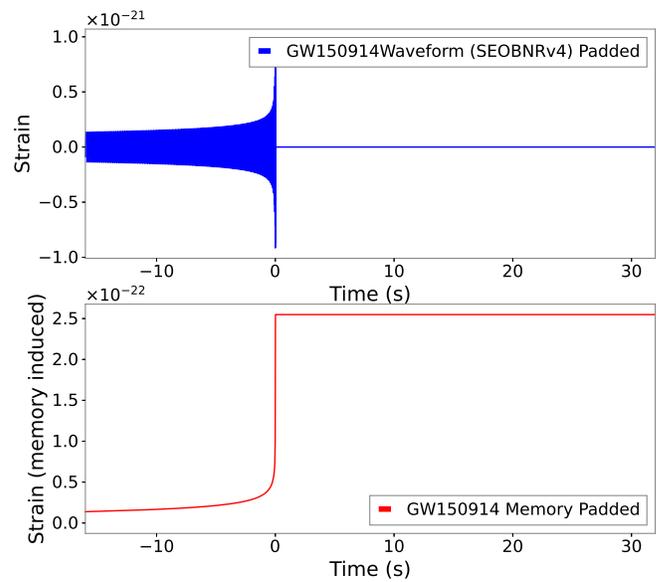}
        \caption{Top panel: GW150914 strain waveform (SEOBNRv4 approximant), Bottom panel: The induced memory strain sourced by the gravitational wave radiation.}
        \label{fig:GW150914memory_only}
\end{figure}

\noindent The first direct detection of gravitational wave (GW) radiation from compact binary sources was made by the LIGO Scientific Collaboration in 2015. This detection and the 89 since validate Einstein's General Relativity in the linearized approximation. The next step in the puzzle is to use the ensemble of detected events to extract salient features predicted by full non-linear GR. In 1991, Christodulou \cite{Christodulou} estimated the strain induced by one particular ramification of non-linear GR, the gravitational memory effect, which predicts a permanent non-zero strain in space-time after the passage of the GW wave. This phenomenon for compact binary sources was estimated to be only one or two orders of magnitude below the GW strain predicted by linearized GR. Shortly after Christoudulou's findings, A. G. Wiseman, C. M. Will, and K. S. Thorne \cite{Clifford} identified that Christodulou's memory effect is essentially sourced by the outgoing radiation itself, i.e. in particle energy language, the additional strain induced by the energy of moving gravitons. Consequently, they derive an expression for the time-dependent strain induced by the memory effect in the Transverse Traceless (TT) gauge, which matches Christodulou's findings in the $t\rightarrow \infty$ asymptotic limit:
\begin{equation}
    h_{\text{Mem}ij}(t)=\frac{4G}{rc^4}\int_{-\infty}^{t}\int_{\Omega'}\frac{dE}{dt d\Omega'}\frac{n'_i n'_j}{1-\overrightarrow{n'}\cdot \overrightarrow{z'}}d\Omega' dt
    \label{eq:memory_energyterms}
\end{equation}
Here $\frac{dE}{dt d\Omega'}$ is the energy flux of the radiated gravitational wave per unit solid angle. The primed coordinates represent the coordinates of the source frame (see Fig.\,\ref{fig:Coordinates_source}), $\overrightarrow{z'}$ being the line of sight vector. Spatial integration is over the entire spherical wavefront centered on the source passing by the detector at time $t$. The time integration represents the addition of these infinitesimal strain contributions from these wavefronts over the entire past right up to the present, thus representing the "memory". Fig.~\ref{fig:GW150914memory_only} shows the strain and the corresponding induced memory strain for GW150914, the first event detected by LIGO detectors \cite{GW150914}, assuming it to be an equal mass binary and using a simplified version of  Eqn.~\ref{eq:memory_energyterms} see Section\,\ref{sec:LISA detector response to Memory waveform}.\\\\
\noindent We see that for stellar and intermediate mass compact binaries, the memory waveform is an order of magnitude weaker than the radiation waveform. This necessitates using an ensemble stack of detected events wherein the cumulative SNR statistic on an average scales with the square root of the number of detections. Grant \emph{et.al} \cite{GWmemmostrecent} have estimated the required detector runtime to detect gravitational memory directly in the data streams of ground-based detectors.\\\\
In this work, we propose an alternative idea of searching for memory imprints in data streams of future space-based GW detectors  (LISA \cite{L3proposal} and follow on LISA-like mission concepts ALIA~\cite{ALIA_BBO}, AMIGO~\cite{AMIGO} and Folkner~\cite{Folkner})using event triggers from ground-based detectors (LIGO, Einstein Telescope (ET)\cite{ET}, and Cosmic Explorer (CE)\cite{CE_horizon,reitze2019cosmic}). One of the advantages of using data streams of space-based GW detectors is the fact that the round trip light travel time is  1 to 4 orders of magnitude greater than the rise time of the memory waveform. Divakarla in his thesis \cite{Atul} showed that the rise time scales as $\mathcal{O}(\frac{GM_{\rm{total}}}{c^3})$ and we calculate the rise time (using the same definition) for the most massive equal mass binary considered in this paper ($M_{\rm{total}}=10^4 M_{\odot}$) to be about $0.5 $\,\rm{sec}, which is a factor of 33 less than the round trip light travel time of LISA \cite{L3proposal}. This difference in time scales results in the memory signal's duration being predominantly determined by the round trip light travel time, the morphology of the signal being determined predominantly determined by the source's sky position (with respect to the detector), and all the other parameters determining only the amplitude of the signal (see Fig.\ref{fig:GW150914memory_armlength_only}). This enables us to use a simple stacking algorithm in time domain and consequently a single template waveform to compute the memory SNR for all events.
\\\\
The paper is organized as follows: Section\,\ref{sec:LISA detector response to Memory waveform} expands on Eqn.\,\ref{eq:memory_energyterms} and calculates the memory waveform response in LISA's TDI X data stream. 
In Section\,\ref{sec:SNR_calculation} we introduce a cumulative SNR statistic formed via the aforementioned stacking or combination of individual memory effects, utilizing GW150914's memory effect as a toy model for estimating the effects of stacked signal on this cumulative SNR. We also use the existing catalog of LIGO events to estimate the LISA detector runtime required to accumulate enough memory SNR to cross the preset threshold of \SNRthreshold. Section\,\ref{sec:Estimating memory SNR using triggers from Cosmic Explorer} scales this calculation to estimate the runtime required if using predicted catalogs from future ground-based detectors triggers like LIGO (with A$^{\#}$ sensitivity), Cosmic Explorer and Einstein Telescope instead of LIGO triggers. Section\,\ref{sec:Prospects of memory detection with future LISA like missions} extends the calculation in Section\,\ref{sec:Estimating memory SNR using triggers from Cosmic Explorer} to future LISA-like mission concepts, namely ALIA, AMIGO, and Folkner. Section\,\ref{sec:Conclusion} discusses conclusions and future implications.

\section{LISA detector response to memory waveform}
\label{sec:LISA detector response to Memory waveform}
\begin{figure}[ht]
        \includegraphics[scale=0.35]{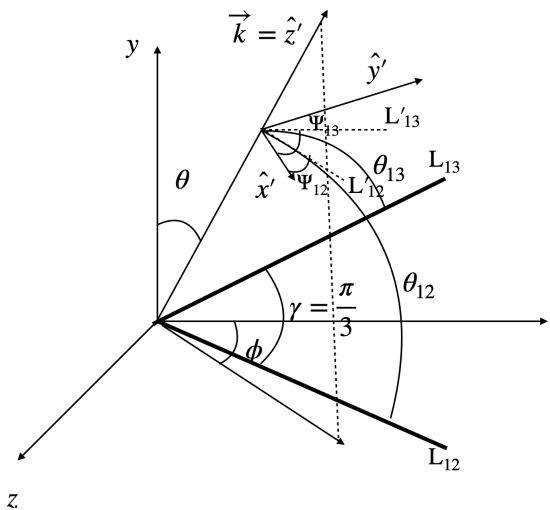}
        \caption{Detector frame Coordinate system. $x',y',z'$ represent the source frame coordinates. $\rm{L}_{12}, \rm{L}_{13}$ represent the LISA arms with and opening angle of $\gamma=\pi/3$, SC$_{1}$ being the primary spacecraft. The detector $x$-axis is fixed at the small angle bisector of the arms while the $z$-axis is fixed to be perpendicular to both arms. The polarization angles $\Psi_{12},\Psi_{13}$  are then defined as the angles made by the LISA arms projected onto the polarization plane ($\rm{L}'_{12},\rm{L}'_{13}$) with the $x'$ axis}
        \label{fig:Coordinates_detector}
\end{figure}

\begin{figure}[ht]
        \includegraphics[scale=0.35]{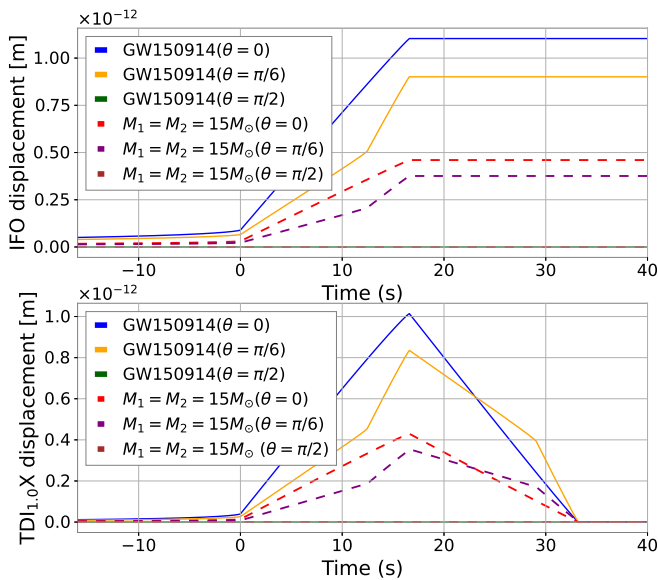}
        \caption{Plots for the GW150914 memory induced test mass (TM) displacement in the LISA IFO (top panel) and TDI$_{1.0}$X (bottom panel) data streams. Traces are produced with the azimuthal angle fixed at $\phi=0$ and values for polarization angles $\Psi_{12}$ and $\Psi_{13}$ that give optimal positively signed memory effect. For comparison, similar traces are plotted for a $15\,M_{\odot}$ equal mass zero spin compact binary situated at the same luminosity distance.}
\label{fig:GW150914memory_armlength_only}
\end{figure}

\noindent The choice of source frame coordinates in Fig.\,\ref{fig:Coordinates_source} ensures that the memory strain waveform is polarized in the $x'-y'$ (+) direction. Furthermore, the memory strain tensor of Eqn.\,\ref{eq:memory_energyterms} can be simplified to an expression involving luminosity distance $\lumdist$, the inclination angle $\iota$ and the largely dominant $l=2,m=2$ spin-weighted spherical harmonic for equal mass non-spinning binary systems \cite{Atul} (Note: All calculations in this paper assume equal mass binaries with zero spin)
\begin{align}
h(t)=&h_{\text{Mem+}}(t)=h_{\text{Mem}11}=-h_{\text{Mem}22}\nonumber\\=&\frac{\lumdist}{192\pi c}\sin^2\iota(17+\cos^2 \iota)\int_{-\infty}^{t}\vert \dot{h}_{22}(t)\vert^2 dt \label{eq:memory_strainterms}\\
h_{\text{Mem}\times}(t)=&h_{\text{Mem}12}=h_{\text{Mem}21}=0
\end{align}
To calculate the memory-induced strain response in LISA arms, we follow \cite{LISA_strain_response} and introduce a new set of "detector frame" coordinates defined in Fig.\,\ref{fig:Coordinates_detector}.
With these conventions,  the strain response ($z_{1j}$) in the LISA arms (equal arms with SC$_{1}$ being the primary spacecraft and arm lengths staying constant over light travel time scales) is given in terms of the memory strain $h(t)$, the inter-SC light travel time $\tau=8.3~\rm{sec}$, the polarization angles $\Psi_{12},\Psi_{13}$ and the angles the arms subtend along the propagation direction $\theta_{12},\theta_{13}$~\cite{LISA_strain_response}:
\begin{widetext}
\begin{equation}
    z_{1j}(t)=\frac{\cos{(2\Psi_{1j})}}{2\tau} \left(\int_{t-2\tau}^{t}h(t)dt-\cos{\theta_{1j}}\left(\int_{t-\tau(1+\cos \theta_{1j}) }^{t}h(t)dt-\int_{t-2\tau}^{t-\tau(1+\cos \theta_{1j})}h(t)dt\right)\right);\,j=2,3\label{eq:Memoryinduced_detectorstrain}
\end{equation}
\end{widetext}
The first integral represents the arm response for waves incoming orthogonal to the detector plane while the second and third integrals represent the corrections for oblique incidence resulting in the wave arriving at slightly different times at every spacecraft.\\\\
LISA's post-processing algorithm, Time Delay Interferometry (TDI) \cite{TDI1} is designed to mitigate the dominant laser frequency noise to the sub picometer level, after which shot noise is expected to dominate above $3\,\text{mHz}$ and acceleration noise dominates below $3\,\text{mHz}$. We assume a shot noise dominated noise background with a white noise displacement amplitude spectral density (ASD) at each science photodiode given by \cite{freqplanning_doc}: 
\begin{equation}
\tilde{n}_s=\frac{5\text{\,pm}}{\sqrt{\text{Hz}}}
\label{eq:Shot_noise_req_per_PD}
\end{equation}
Thus the noise contribution to the single arm response is
\begin{equation}
    n_{1j}(t)=n_{s1j}(t)-n_{sj1}(t-2\tau);~ j=2,3
\end{equation}
where $\tau$ is the one way light travel time and $n_{sij}$ is the shot noise on the science photodiode on spacecraft $i$ which is linked to spacecraft $j$. The total single arm and the interferometer strain responses are given by:
\begin{align}
    s_{1j}(t)=&z_{1j}(t)+n_{1j}(t);~j=2,3\\
    \Delta_{1}(t)=&s_{12}(t)-s_{13}(t)
\end{align}
For this paper, we use the TDI$_{1.0}$ $X$ data stream, which results from the first-generation TDI algorithm that assumes a constant arm length over the time scale of the round trip light travel time. Additionally for computational simplicity, we also assume equal interferometer arms. With this assumption the TDI$_{1.0}$ $X$~\cite{TDI1} response is simplified to:
\begin{align}
X_{1.0}(t)=&\Delta_{1}(t)-\Delta_{1}(t-2\tau)
\end{align}
Note that for LISA's science operations, it is pertinent to account for unequal arms to subtract out laser frequency noise. Given that LISA's arm length mismatch corresponds to a light travel time $\Delta \tau<130\,\text{msec}$ we estimate the resulting correction to the memory SNR computed with the equal arm assumption to be smaller by a factor of $O(10^{-3})$. Furthermore, LISA's science operation will most likely employ the second-generation TDI algorithm TDI\,${2.0}$ which accounts for changes in arm length during one round trip light travel time. While we do not expect significant (if any) changes in the memory SNR computed with the TDI$_{1.0}\,X$ data stream, we aim to use TDI\,${2.0}$ data streams in future works.\\\\
Fig.\,\ref{fig:GW150914memory_armlength_only} illustrates the length change induced by the GW 150914 memory in the LISA interferometer and TDI$_{1.0}$ $X$ responses, for a fixed azimuthal angle $\phi$ with different polar angles ($\theta$ in Fig.\,\ref{fig:Coordinates_detector}) and corresponding optimal values (with respect to the interferometer (IFO) and TDI responses) of the polarization angles $\Psi_{12}$ and $\Psi_{13}$. Note that the choice of polarization angles chosen for the figure are optimal for the positively signed memory effect. The transformation $\left(\Psi_{12},\Psi_{13}\right)\rightarrow \left(\frac{\pi}{2}+\Psi_{12},\frac{\pi}{2}+\Psi_{13}\right)$ gives the corresponding negatively signed memory effect.

\begin{figure}[ht]
        \includegraphics[scale=0.35]{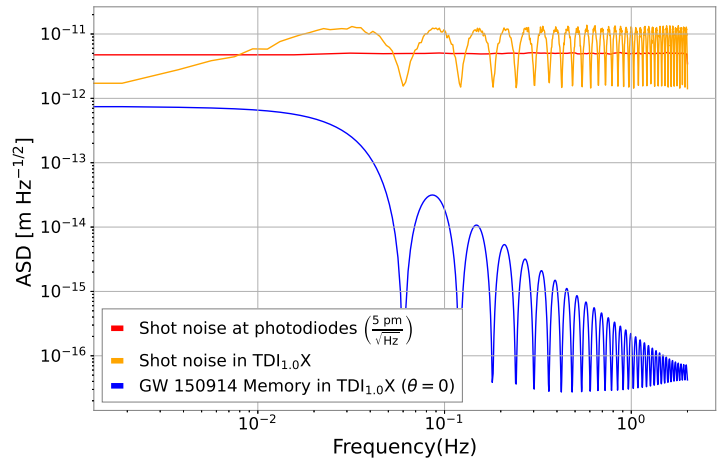}
        \caption{GW 150914 memory (optimal orientation) against shot noise background in TDI$_{1.0}$X}
\label{fig:Memoryvsshotnoise_frequencydomain}
\end{figure}

\section{Memory SNR Calculation}
\label{sec:SNR_calculation}
\noindent As evident from Fig.\,\ref{fig:Memoryvsshotnoise_frequencydomain}, shot noise dominates the GW150914 response and the same is true for other stellar/intermediate mass compact binaries. Furthermore, from Eqn.\,\ref{eq:Memoryinduced_detectorstrain} and Fig.\,\ref{fig:GW150914memory_armlength_only}, it is evident that the memory response in TDI X resembles 
(further quantified in Section\,\ref{subsec:Memory SNR of GW 150914 like events in LISA}) a triangular pulse ($h_{\Delta}(t)$) with a base of $4\tau$ starting from the coalescence time. These two observations motivate the idea of looking up the coalescence times of recorded GW events and creating a composite data stream, $X_{\rm{stack}}(t)$, by stacking up the TDI data snippets $(X_{i})$ of length $4\tau~\rm{sec}$ that start at these coalescence times. This stacking is performed with the appropriate weight factors $(w_{i})$ and is processed through a common-matched filter with a triangular pulse template. (Note: Excluding the effects of sub-optimal inclination angle and  sky direction angles (see Section\,\ref{subsec:Memory SNR of GW 150914 like events in LISA}), the use of a triangular pulse template as opposed to the template derived from the memory waveform itself results in a $<0.2\%$ loss in SNR in the worst-case scenario, the worst-case scenario being a binary with $M_{\text{Tot}}=10^4 M_{\odot}$). \\\\
We divide this section into three subsections. In Section\,\ref{subsec:The weighted SNR statistic}, we define the cumulative SNR statistic and define the corresponding detection threshold SNR value used for the extraction of memory signals. Following this in Section\,\ref{subsec:Memory SNR of GW 150914 like events in LISA}, we apply our SNR statistic on a stack of GW 150914 like events. Accounting for expected loss of SNR due to sub-optimal detector orientation (orange and green traces in Fig.\,\ref{fig:GW150914memory_armlength_only}) and losses in the source power due to  source inclination angle (Eqn.\,\ref{eq:memory_strainterms}), we calculate an estimate of the number of GW 150914-like events required to cross the SNR threshold. Finally in Section\,\ref{subsec:Memory SNR of the LIGO O3 Catalog}, we use scaling arguments to calculate the expected memory SNR in LISA accumulated with events detected during LIGO's O3 science run \cite{GWTC2,gwtc3}. We estimate the expected time required for LIGO-like detectors (with O3 sensitivity) to provide enough detections for the cumulative memory SNR in LISA to cross the preset threshold.

\subsection{The memory SNR statistic}
\label{subsec:The weighted SNR statistic}
\noindent In this paper, the definition of SNR of a generic data stream $d(t)$ matched to a triangular pulse template of unit height $h_{\Delta}(t)$ is given by
\begin{equation}
    \text{SNR}=\frac{\overrightarrow{d(t)}^{\dag}C^{-1}\overrightarrow{h_{\Delta}(t)}}{\sqrt{\overrightarrow{h_{\Delta}}^{\dag}C^{-1}\overrightarrow{h_{\Delta}(t)}}}
    \label{eq:SNR_time_domain_definition}
\end{equation}
where $C$ is the noise covariance matrix of $d(t)$ whose matrix elements are given by:
\begin{equation}
    C_{ij}=\langle\left(\text{noise}(d[i])^*\right) \left(\text{noise}(d[j])\right)\rangle.
\end{equation}
where $d[i],d[j]$ represent the $i^{\rm{th}}$ and $j^{\rm{th}}$ elements of the data stream $d(t)$ and the angled brackets represents an ensemble average.
This definition of SNR is equivalent to the well-known frequency domain definition of the optimum matched filter SNR used in LIGO's modeled search analysis \cite{DuncanBrown}. For a zero-mean noise background, we see that in the absence of a signal, the SNR statistic in Eqn.\,\ref{eq:SNR_time_domain_definition} has zero mean and standard deviation ($\sigma_{\text{SNR}}$) of unity. This motivates us to use an SNR threshold of $\SNRthreshold$ to claim a $\SNRthreshold\sigma$ detection.\\\\
As mentioned above, the data stream for our calculation is a weighted combination of $N$ TDI data snippets with a background dominated by shot noise. Therefore, the data stream is given by
\begin{equation}
    \overrightarrow{d(t)}=\sum_{i=1}^{N}w_{i}\overrightarrow{X_{i}(t)}
\end{equation}
and the covariance matrix takes the form:
\begin{align}
C=&\left(\sum_{i=1}^{N}w_i^2\right)\langle n_{\rm shot}^2(t)\rangle\begin{pmatrix}8\rm{I}_{2\tau\times 2\tau}&-4\rm{I}_{2\tau\times 2\tau}\\
    -4\rm{I}_{2\tau\times 2\tau}&8\rm{I}_{2\tau\times 2\tau}\\
\end{pmatrix}\\
    =&\left(\sum_{i=1}^{N}w_i^2\right)C_{\text{1 event}}
\end{align}
Here $\langle n^2_{\rm shot}(t)\rangle=\tilde{n}_s^2\frac{f_{s_{\rm{LISA}}}}{2}$ is the variance of the shot noise in each science photodetector (derived from Eqn.\,\ref{eq:Shot_noise_req_per_PD}), LISA's sample rate being $f_{s_{\rm{LISA}}}=4~\rm{Hz}$ and the highest frequency in band being given by the Nyquist criterion to be 2 Hz. Additionally, $I_{2\tau\times 2\tau}$ represents a $66 \times 66 $ identity matrix $\left(66=\text{Greatest Integer}(2\tau f_{s_{\rm{LISA}}})\right)$. \\

The optimum choice of weights $w_i$ corresponds to the choice that maximizes the expected value of the SNR statistic in the presence of a signal.
Modeling the signal component of $X_{i}$ to be a triangle of known height $a
_i$, we have 
\begin{equation}
    \text{signal}(X_{i}(t))\approx a_{i} h_{\Delta}(t).
\end{equation}
Therefore, using Eqn.\,\ref{eq:SNR_time_domain_definition}, the expected value of SNR in presence of a signal is given by 
\begin{equation}
    \langle \text{SNR}\rangle=\left(\frac{\sum_{i=1}^{N}a_{i}w_{i}}{\sqrt{\sum_{j=1}^{N}w_{j}^2}}\right)\sqrt{h_{\Delta} C^{-1}_{\text{1event}}h_{\Delta}},
\end{equation}
which has a maximum value of
\begin{equation}
        \langle\text{SNR}\rangle_{\text{Max}}=\sqrt{\left(\sum_{i=1}^{N}a_{i}^2\right)}\sqrt{h_{\Delta} C^{-1}_{\text{1event}}h_{\Delta}}
\label{eq:SNR_max_expectation}        
\end{equation} 
for the choice of weights $w_i=a_i$.
\\

Note that here we make the critical assumption of knowing the sign of the memory signal (i.e., the sign of each $a_i$), which cannot be determined by Eqn\,\ref{eq:memory_strainterms} alone. Lasky et al.\,\cite{Lasky_2016} propose a method to deduce the sign of the memory effect from the measured higher-order modes of the oscillating GW waveform.
\subsection{Memory SNR of GW 150914 like events in LISA}
\label{subsec:Memory SNR of GW 150914 like events in LISA}
\begin{figure}[ht]
        \includegraphics[scale=0.35]{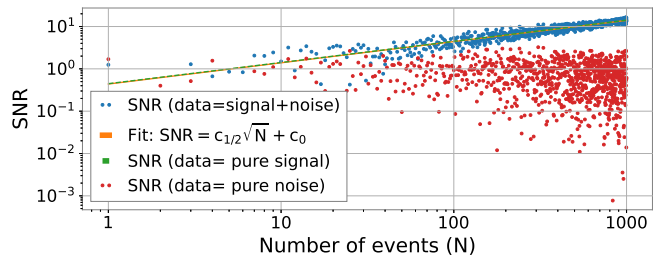}
        \caption{Scaling of matched filter SNR with number of optimally oriented GW150914 events.}
        \label{fig:SNR_scaling_matched_filter}
\end{figure}

Using the definition of SNR in Section\,\ref{subsec:The weighted SNR statistic} we calculate the memory SNR accumulated by stacking GW150914-like events. We simulate multiple TDI X data streams each composed of the GW150914 memory signal with optimal source inclination angle ($\iota=\pi/2$) and optimal detector orientation ($\theta=0$ in Fig.\,\ref{fig:Coordinates_detector}) in a randomly generated shot noise background. Fig.\,\ref{fig:SNR_scaling_matched_filter} shows the accumulation of SNR as a function of number of stacked event data streams. We see that the signal rises above the noise background on stacking about 5 data streams and eventually end up with an SNR of \SNRthousand\, on stacking 1000 data streams. The figure also confirms the square-root dependence of the SNR on the number of events. From this, we can extrapolate and estimate the average SNR of one optimally oriented GW 150914 event to be $0.45$, and therefore require a stack of about 125 optimally oriented GW150914-like events to cross the SNR threshold of 5.
\\\\
We now use Eqns.~\ref{eq:memory_strainterms} and \ref{eq:Memoryinduced_detectorstrain} to estimate the effects of averaging over inclination angle $\iota$ and the sky-direction angles $(\theta,\phi)$, i.e., to estimate the SNR lost from the diminished signal strength of sub-optimally oriented events and the SNR lost from using the triangular template for extracting  these sub-optimally oriented events. These three angles are assumed to be mutually independent and uniformly distributed.
We use Eqn.\,\ref{eq:memory_strainterms} estimate the effects of averaging over inclination angle $\iota$ and  Eqn.\,\ref{eq:Memoryinduced_detectorstrain} to estimate the effects of averaging over the sky-direction $(\theta,\phi)$ and polarization $(\Psi_{12},\Psi_{13})$ angles. The four angles $\iota$,$\theta$, $\phi$ and $\Psi_{+}=\Psi_{12}+\Psi_{13}$ are mutually independent and are assumed to be uniformly distributed, while $\Psi_{-}=\Psi_{12}-\Psi_{13}$ is a function of $\theta$ and $\phi$.\\\\
The ratio of the SNR averaged by the inclination angle and the maximum SNR is given by
\begin{equation}
    \frac{\langle\text{SNR}(\iota)\rangle_{\iota}}{\text{SNR}(\iota=\pi/2)}=\frac{\langle\sin^2\iota(17+\cos^2 \iota)\rangle}{17}\approx\SNRincavg
    \label{eq:SNR_avg_inclination} 
\end{equation}

\begin{figure}[ht]
        \includegraphics[scale=0.35]{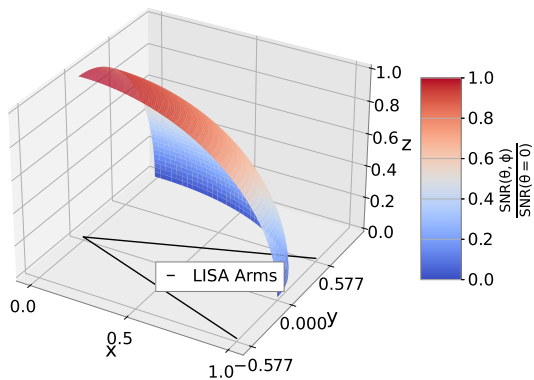}
        \caption{Variation of SNR with sky-direction with optimal polarization angles ($\Psi_{+}=\pi/2$).}
        \label{fig:Average_memory_skydirection}
\end{figure}

\noindent Similarly, we use Eqn.\,\ref{eq:Memoryinduced_detectorstrain} to estimate the fractional drop in SNR as we move away from orthogonal incidence to the plane of the detector (see Fig.\,\ref{fig:Average_memory_skydirection}). Consequently, average over sky-directions and polarization angles as a fraction of the maximum SNR$(\theta=0,\Psi_{+}=\pi/2)$
\begin{equation}
    \frac{\langle\text{SNR}\rangle_{\theta,\phi,\Psi_{+}}}{\text{SNR}(\theta=0,\Psi_{+}=\pi/2)}=0.55\frac{2}{\pi}
    \label{eq:SNR_avg_theta_phi}
\end{equation}

\noindent From Eqns.\,\ref{eq:SNR_avg_inclination} and \ref{eq:SNR_avg_theta_phi}, we deduce the averaged memory SNR from one GW150914-like event to be $\SNRoneeventtotavg\left(=\left[(0.5)\left(0.55\frac{2}{\pi}\right)\right](0.45)\right)$. Thus, approximately $\ROUNDnumeventsSNRthreshold$ randomly oriented GW150914-like events are needed to cross an SNR threshold of \SNRthreshold.
\subsection{Memory SNR of the LIGO O3 Catalog}
\label{subsec:Memory SNR of the LIGO O3 Catalog}
\begin{figure}[ht]
        \includegraphics[scale=0.35]{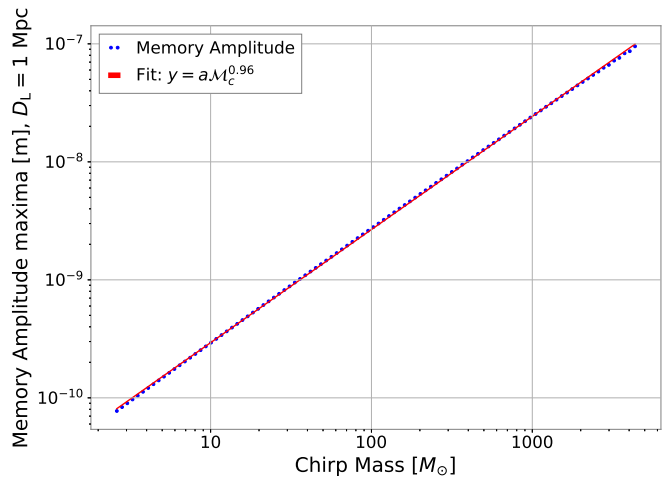}
        \caption{Memory waveform maxima scaling with chirp mass $\mathcal{M}_c$ for binaries located at a luminosity distance $\lumdist=1~\text{Mpc}$}
        \label{fig:Memory_amplitude_chirp_mass_scaling}
\end{figure}

\noindent To get a rough estimate of the memory SNR from the LIGO O3 catalog events, we estimate the scaling relation between the memory amplitude ($a$) and the binary chirp mass ($\mathcal{M}_{c}$) for the SEOBNRv4 waveform model class in Fig.\,\ref{fig:Memory_amplitude_chirp_mass_scaling}.\\

\noindent From the graph, we imply an empirical scaling relation:
\begin{equation}
    a\propto \lumdist^{-1}(\mathcal{M}_c)^\alpha; \quad \alpha=0.96
    \label{eq:memory_scaling_with_mass}
\end{equation}
(Note that the $\lumdist^{-1}$ dependence comes from Eqn.\,\ref{eq:memory_strainterms}). 
Furthermore, for the events in the LIGO catalog with O3 sensitivity \cite{gwtc3,GWTC2}, we see that
\begin{align}
    \sqrt{\sum_{\rm{Catalog Events}}\lumdist^{-2}(\mathcal{M}_c)^{2\alpha}} &\gtrapprox \numeventsperligocatalog(\lumdist^{-1}(\mathcal{M}_c)^\alpha)_{150914}\\
    \therefore \sqrt{\sum_{\rm{Catalog Events}}a^2} &\gtrapprox \numeventsperligocatalog \, a_{150914}
\end{align}
and consequently, from Eqn.\,\ref{eq:SNR_max_expectation} the cumulative memory SNR of the O3 catalog is related to the memory SNR of GW 150914 by:
\begin{align}
    \rm{SNR}_{\rm{O3 Catalog}}\gtrapprox& \numeventsperligocatalog\, \rm{SNR}_{150914}=\SNRoneLIGOthreecatalog,
\end{align}
which means that the LIGO O3 run will have be repeated at most \ROUNDnummrepeatligocatalog\,times over to cross the SNR threshold of \SNRthreshold.\\\\
From \cite{O1O2,gwtc3,GWTC2},  we find the runtime for LIGO's O3 run to be $t_{\rm{LIGO-O3} }\approx \timeyrsligocatalog~\rm{yr}$. Thus for LIGO detectors with O3 sensitivity, the cumulative memory SNR in LISA would cross the detection threshold in $\approx \ROUNDtimeyrsligocatalogforSNRthreshold~\rm{yrs}$ of concurrent operation.\\\\
That being said, we do expect to have far more sensitive ground-based detectors like LIGO with A$^{\#}$ sensitivity, Cosmic Explorer, and Einstein Telescope to be up and running when LISA starts its science operation. From the horizon redshift curves plotted in Fig.\,\ref{fig:horizon_redshift} we expect the event detection rate from these detectors to be much higher than LIGO with O3 sensitivity thus resulting in a much faster memory SNR accumulation in LISA.
Section\,\ref{sec:Estimating memory SNR using triggers from Cosmic Explorer}  provides estimates for the required runtime to detect gravitational memory in LISA from the projected catalogs of these future ground-based detectors.\\\\ 
Note on the error analysis: In this section, we estimated values for the required number of GW150914-like events, the number of LIGO O3 catalogs, and consequently, the time required for LISA (using the LIGO O3 catalog) to accumulate a memory SNR of \SNRthreshold. Assuming perfect knowledge of the waveform parameters, the uncertainty in the values is solely attributed to the uncertainty in the SNR statistic (Eqn.\ref{eq:SNR_time_domain_definition}). The SNR statistic has a unit variance, and thus, the upper and lower bounds $1\,\sigma$ error bar for the aforementioned quantities corresponds to expected values for SNR thresholds of \ERRORBARUPSNRthreshold \, and \ERRORBARDOWNSNRthreshold\,respectively. These quantities have a square root dependence on the SNR, which results in the upper and lower $1\, \sigma$ error bounds being $21\%$ above and $19\%$ below the mean value respectively. Assuming perfect knowledge of the astrophysical event rates and source distributions, the  same error bar applies for values of required detector runtime quoted in Tables\,\ref{tab:SNR_accumulation} and \ref{tab:SNR_accumulation_beyondLISA}.\\\\

\section{Estimating memory SNR in LISA using triggers from future ground-based detectors}
\label{sec:Estimating memory SNR using triggers from Cosmic Explorer}
\begin{figure}[ht]
    \centering
    \includegraphics[scale=0.35]{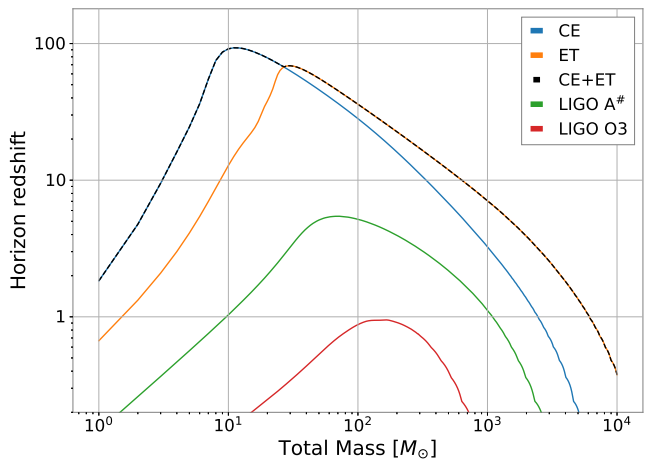}
    \caption{Horizon redshifts of ground-based detectors \cite{Horizon_redshift}. The traces represent the maximum redshift at which optimally oriented equal mass binaries can be detected with an SNR threshold of 8.}
    \label{fig:horizon_redshift}
\end{figure}

\noindent In this section we estimate the expected rate of accumulation of memory SNR corresponding to triggers from future ground-based GW detectors namely, Advanced LIGO with A$^{\#}$ sensitivity, Einstein Telescope \cite{ET} and Cosmic Explorer \cite{CE_horizon,reitze2019cosmic} by scaling the result from the LIGO's O3 observing run. The cumulative memory SNR corresponding to triggers from a particular ground-based detector (d) with a runtime of $T_{\rm{obs}}\,[\rm{yrs}]$ is expressed as :
\begin{widetext}
\begin{equation}
    \text{SNR}^{\rm{d}}=\kappa \sqrt{T_{\rm{obs}}} \sqrt{\int_{M_{\text{min}}}^{M_{\text{max}}} \int_{z_{\text{min}}} ^{z^{\rm{d}}_{\text{max}}(M)} \left[\frac{M^{\alpha}(1+z)^{\alpha}}{\lumdist(z)}\right]^2 \left[\sum_{\rm{source:s}}\beta_s P_s(M)R_s(z)\right] \left[4 \pi (\comdist(z))^2 \frac{d \comdist(z)}{dz} dz\right] dM}
\label{eq:SNR_memory_full_scaling}    
\end{equation}
\end{widetext}
The double integral over redshift and mass essentially represents the expected sum of squared heights of the memory waveforms corresponding to all the triggers provided by detector d in one year of observation. From Eqn.\,\ref{eq:SNR_max_expectation} we see that the SNR has a square-root dependence on this sum. For the purposes of this paper, we have chosen to restrict the calculation to equal mass binaries with the total mass ranging from $1~M_{\odot}-10^4 M_{\odot}$. The corresponding upper limit for the redshift integral is therefore given by the ground-based detector's redshift horizon $z^{\text{d}}_{\text{Max}}(M)$ i.e., the maximum redshift at which an equal mass binary to total mass $M$ can be detected with an SNR of 8. These redshift horizon curves are computed (and plotted in Fig.\,\ref{fig:horizon_redshift}) for Advanced LIGO with O3 and A$^{\#}$ sensitivities, Einstein Telescope, and Cosmic explorer using the GitHub code referenced in \cite{Horizon_redshift}. Thus, the domain of integration is represented by the area under the horizon redshift curves (albeit for computational purposes of circumventing singularities in the luminosity distance expression at $z=0$, we employ a lower redshift cutoff of $z_{\text{min}}=0.01$ which is the closest source that the LVK has measured to date).\\\\
The integrand (which is independent of the ground-based detector's sensitivity) is a product of the following terms:
The first term, $\frac{M^{\alpha}(1+z)^{\alpha}}{\lumdist(z)}$, is essentially proportional to the height of the memory contribution of a single event with total mass $M$ at a luminosity distance $\lumdist(z)$(from Eqn.\,\ref{eq:memory_scaling_with_mass}).
The second term, $\sum_{\rm{source:s}} \beta_s P_s(M)R_s(z)$, is the rate density per unit comoving volume per unit time at which binaries of total mass M occur $\left(\text{units of} \frac{\text{Number}}{M_{\odot} \rm{Gpc^3} \rm{yr}}  \right)$. $\beta_s$ represents the relative abundance of a formation channel, $P_{s}(M)$ represents the normalized Initial Mass Function (IMF) for that channel, and $R_{s}(z)$ represents the corresponding rate. In this paper, we have assumed Globular Clusters (GC) and Isolated Field Binaries (IF)  as the dominant formation channels with the relative abundance of $\beta_{\text{GC}}=\beta_{\text{IF}}=0.5$\,\cite{Ng_2021}. The IMF for field binaries, $P_{\text{IF}} $ is assumed to be a power law distribution \cite{IMF_IF} while the one for Globular Clusters, $P_{\text{GC}}$, is assumed to be log-uniform \cite{IMF_GC}. The corresponding rates as a function of redshift have been taken from \cite{CE_horizon,Ng_2021}.
The third term in the integrand, $4 \pi (\comdist(z))^2 \frac{d \comdist(z)}{dz}$, represents the infinitesimal comoving volume element, with $\comdist(z)$ representing the radial comoving distance.
The cosmological model that gives the expressions for $\lumdist(z)$ and $\comdist(z)$ is assumed to be the LCDM cosmology derived by the Planck Collaboration \cite{LCDM_cosmology}.
\\\\
Having defined all the terms in Eqn.\,\ref{eq:SNR_memory_full_scaling}, Table \ref{tab:SNR_accumulation} shows the ratio of the rate of memory SNR accumulation in LISA by using projected  LIGO A$^{\#}$, ET and CE catalogs with the SNR accumulation in LISA  using the existing LIGO O3 catalog as a baseline. Additionally, an estimate for the run time required for a $\text{SNR}=\SNRthreshold$ detection is also computed. The results suggest that memory detection with LISA, given a lifetime of 10 yrs, is possible only if we use ET or ET in combination with CE. This comes from ET having the larger horizon redshift for high mass binaries compared to other ground-based detectors as shown in Fig.\,\ref{fig:horizon_redshift}.\\\\
\begin{table}[ht]
\begin{center}
\scalebox{0.85}{
    \begin{tabular}{|c|c|c|}
    \hline
        Detector(d) & $\frac{\text{SNR}^{d}}{\text{SNR}^{\text{LIGO O3}}}$ & Time for memory detection with LISA[yrs]\\
        \hline
                CE &\SNRaccumationrateCE&$\ROUNDtimeyrsCEforSNRthreshold$\\
        \hline
                        ET  &\SNRaccumationrateET&$\ROUNDtimeyrsETforSNRthreshold$\\
        \hline
                        LIGO A$^{\#}$  &\SNRaccumationrateligoAsharp&$\ROUNDtimeyrsligoAsharpforSNRthreshold$\\
        \hline
                        LIGO O3  &\SNRaccumationrateLIGO&$\ROUNDtimeyrsligocatalogforSNRthreshold$\\
        \hline
           
                CE+ET &\SNRaccumationrateCEandET&$\ROUNDtimeyrsCEandETforSNRthreshold$\\
                \hline
    \end{tabular}}
\caption{Time taken to accumulate an SNR threshold of $\SNRthreshold$ in LISA's TDI data stream using triggers from projected CE, ET and LIGO A$^{\#}$ catalogs.}
\label{tab:SNR_accumulation}
\end{center}
\end{table}

\section{Prospects of memory detection with future LISA like missions}
\label{sec:Prospects of memory detection with future LISA like missions}
\noindent In addition to LISA we expect some more LISA-like  mission concepts to be up and running after LISA's science operation in the second half of the century, some examples being ALIA, AMIGO, and Folkner. The corresponding results for these missions are presented in Table \ref{tab:SNR_accumulation_beyondLISA}. In this section, we take a look at each one of them briefly.\\\\
AMIGO's white paper document \cite{AMIGO} entails a configuration identical to LISA (with 2.5$\,\text{Gm}$ arms and the same measurement band) but predicts a 10-fold reduction in acceleration noise and an approximate 10-fold reduction in shot noise achieved by using a 30 W laser in conjunction with a slightly bigger telescope diameter of $0.5\,\text{m}$. This in turn leads to an approximate 10-fold improvement in the memory SNR and a 100-fold improvement in the time required to detect the memory effect.\\\\
ALIA's design \cite{ALIA_BBO} entails a equilateral triangle with 0.5$\, \text{Gm}$ arms leading to a slightly higher  measurement 
 band. Additionally, there is a projected  40000-fold increase in received power which leads to a 200-fold improvement of the shot noise limited displacement sensitivity which combined with the arm length reduction results in a  40-fold increase in the shot noise limited strain sensitivity, $\frac{40}{\sqrt{5}}$-fold increase in the memory SNR and consequently a 320-fold improvement in the time required for memory detection.\\\\
In contrast to ALIA and AMIGO, Folkner's interferometer configuration \cite{Folkner} entails  a 260$\,\text{Gm}$ equilateral triangle, thus targeting  frequencies below the LISA band. In the shot noise limit, the increase in armlength is compensated by the reduced power received at the photodetector thus leading to an almost identical strain sensitivity (a factor of ~$\sqrt{2/3}$ improvement coming from using a 3W transmit laser instead of a LISA's 2 W laser). One challenge in using the shot noise dominated limit for Folkner is the fact that most of its frequency band is dominated by the Galactic confusion noise the resolution of which is unclear to date. That being said, assuming the galactic confusion noise is resolvable, we gain memory SNR primarily due to the increased integration time over a round trip (factor of $10.2\sqrt{3/2}$ in SNR and 156 in the time required for detection).

\begin{widetext}
\begin{table}[ht]
\begin{center}
\renewcommand{\arraystretch}{1.5}
\scalebox{0.9}{
    \begin{tabular}{|c|c|c|c|c|c|}
    \hline
        Detector(d) & $\frac{\text{SNR}^{d}}{\text{SNR}^{\text{LIGO O3}}}$ & \multicolumn{4}{c|}{Time for memory detection[yrs]}\\
        \hline
        &&LISA& AMIGO & ALIA& Folkner\\
        \hline
                CE &\SNRaccumationrateCE&$\ROUNDtimeyrsCEforSNRthreshold$&$\ROUNDtimeyrsCEforSNRthresholdAMIGO$&$\ROUNDtimeyrsCEforSNRthresholdALIA$&$\ROUNDtimeyrsCEforSNRthresholdFolkner$\\
        \hline
                        ET  &\SNRaccumationrateET&$\ROUNDtimeyrsETforSNRthreshold$&$\ROUNDtimeyrsETforSNRthresholdAMIGO$ &$\ROUNDtimeyrsETforSNRthresholdALIA$&$\ROUNDtimeyrsETforSNRthresholdFolkner$\\
        \hline
                        LIGO A$^{\#}$  &\SNRaccumationrateligoAsharp&$\ROUNDtimeyrsligoAsharpforSNRthreshold$&$\ROUNDtimeyrsligoAsharpforSNRthresholdAMIGO$&$\ROUNDtimeyrsligoAsharpforSNRthresholdALIA$&$\ROUNDtimeyrsligoAsharpforSNRthresholdFolkner$\\
                \hline
                        LIGO O3  &\SNRaccumationrateLIGO&$\ROUNDtimeyrsligocatalogforSNRthreshold$&$\ROUNDtimeyrsligocatalogforSNRthresholdAMIGO$&$\ROUNDtimeyrsligocatalogforSNRthresholdALIA$&$\ROUNDtimeyrsligocatalogforSNRthresholdFolkner$\\
        \hline
        CE+ET &\SNRaccumationrateCEandET&$\ROUNDtimeyrsCEandETforSNRthreshold$&$\ROUNDtimeyrsCEandETforSNRthresholdAMIGO$&$\ROUNDtimeyrsCEandETforSNRthresholdALIA$&$\ROUNDtimeyrsCEandETforSNRthresholdFolkner$\\
        \hline
    \end{tabular}}
\caption{Time taken to accumulate a SNR threshold of $\SNRthreshold$ in TDI data streams of LISA, ALIA, AMIGO and Folkner using triggers from projected CE, ET and LIGO A$^{\#}$ catalogs}
\label{tab:SNR_accumulation_beyondLISA}
\end{center}
\end{table}

\end{widetext}

\section{Conclusion and Future prospects}
\label{sec:Conclusion}
\noindent In this article we propose a simple data analysis procedure to detect Chirstodulou's non linear gravitational memory effect using LISA's (and future LISA like mission's) data stream and  the event catalog of ground-based detectors. The idea is to first use the detection catalog of ground-based to get arrival times of detected events and then stack up with optimumal weight factors, data snippets from the TDI data stream that start at these arrival times. This is followed up by computing the SNR in time domain by running the composite data stream through a matched filter with a triangular pulse being the template. Using simple astrophysical population models of binary black holes (Section\,\ref{sec:Estimating memory SNR using triggers from Cosmic Explorer}) and the memory SNR derived from the existing LIGO O3 catalog, Table\,\ref{tab:SNR_accumulation_beyondLISA} provides estimates for the time required for the cumulative memory SNR in LISA and future space-based missions like AMIGO,  ALIA and Folkner to cross the SNR threshold of $\SNRthreshold$ for catalogs derived from CE, ET, LIGO with A$^{\#}$ and O3 sensitivities. Results suggest that LISA in conjuntion with ET will most likely be able to detect the gravitational memory effect in few years of concurrent operation and the results for the beyond LISA missions look even more promising. \\\\
While we do believe that the work done provides good baseline estimates and a time domain data analysis method to detect gravitational memory in space-based detectors, there are a few refinements that one can look into. Most of these involve relaxing assumptions to make the estimate more accurate.
\begin{enumerate}
    \item Extending the calculation for unequal mass case. This would involve having to calculate and include higher order spherical harmonic contributions \cite{Higher_modes} in Eqn.\,\ref{eq:memory_strainterms} and having to modify Eqn.\,\ref{eq:SNR_memory_full_scaling}  to a integral over both mass components.
    \item We could avoid the SNR loss due to the $\theta ,\phi $ orientation angles by splitting the sky in sections and assigning a template for each sky section instead of having a universal triangular pulse template. Waveforms corresponding to oblique incidence angles resemble kinked triangular pulses with reduced heights (e.g., orange trace in Fig.~\ref{fig:GW150914memory_armlength_only}). Thus in addition to using different templates for each sky section, we would also need to update the corresponding optimum weight factors used for stacking the data streams.
    \item While in this work we look for memory signals in the TDI$_{1.0}$ $X$ data stream, eventually in will be useful to extend the calculations with TDI$_{2.0}$ data streams accounting for varying armlengths over one round trip light travel time. Intuitively, we do not expect drastic changes from the estimates provided here. 
\end{enumerate}

\section{Acknowledgement}
We would like to thank Bernard Whiting for fruitful discussions. Additionally, this work is supported by NASA (National Aeronautics and Space Administration) grants 80NSSC20K0126 and 80NSSC19K0324 for which we would like to thank NASA for funding the reasearch. Finally,  this research uses the publicly available data from the LVK collaboration's observing runs and we acknowledge the following data statement, "This research has made use of data or software obtained from the Gravitational Wave Open Science Center (gw-openscience.org), a service of LIGO Laboratory, the LIGO Scientific Collaboration, the Virgo Collaboration, and KAGRA. LIGO Laboratory and Advanced LIGO are funded by the United States National Science Foundation (NSF) as well as the Science and Technology Facilities Council (STFC) of the United Kingdom, the Max-Planck-Society (MPS), and the State of Niedersachsen/Germany for support of the construction of Advanced LIGO and construction and operation of the GEO600 detector. Additional support for Advanced LIGO was provided by the Australian Research Council. Virgo is funded, through the European Gravitational Observatory (EGO), by the French Centre National de Recherche Scientifique (CNRS), the Italian Istituto Nazionale di Fisica Nucleare (INFN) and the Dutch Nikhef, with contributions by institutions from Belgium, Germany, Greece, Hungary, Ireland, Japan, Monaco, Poland, Portugal, Spain. The construction and operation of KAGRA are funded by Ministry of Education, Culture, Sports, Science and Technology (MEXT), and Japan Society for the Promotion of Science (JSPS), National Research Foundation (NRF) and Ministry of Science and ICT (MSIT) in Korea, Academia Sinica (AS) and the Ministry of Science and Technology (MoST) in Taiwan."

\

\nocite{*}
\bibliography{GW_memory}

\end{document}